\documentstyle[prl,aps,preprint,eqsecnum]{revtex}
 
\begin{document}
\draft

\def\nid{\noindent}
\def\half{\mbox{\small $\frac{1}{2}$}}
\def\quarter{\mbox{\small $\frac{1}{4}$}}
\def\sfrac#1#2{\mbox{\small $\frac{#1}{#2}$}}
\def\beq{\begin{equation}}
\def\eeq{\end{equation}}
\def\eeql#1{\label{#1} \end{equation}}
\def\bea{\begin{eqnarray}}
\def\eea{\end{eqnarray}}
\def\eeal#1{\label{#1} \end{eqnarray}}


\title{Ambiguities in Quantizing a Classical System}

\author{Ian H. Redmount}

\address{Department of Physics,
Parks College of Engineering and Aviation,
Saint Louis University,\\
PO Box 56907,
St. Louis, Missouri  63156-0907, USA}

\author{Wai-Mo Suen}

\address{McDonnell Center for Space Sciences, Department of Physics,
Washington University, \\
St. Louis, Missouri 63130, USA, and \\
Physics Department,
The Chinese University of Hong Kong, 
Hong Kong, China}

\author{Kenneth Young}

\address{Physics Department,
The Chinese University of Hong Kong, 
Hong Kong, China}

\date{\today}

\maketitle


\begin{abstract}

One classical theory, as determined by an equation of motion or set of
classical trajectories, can correspond to many unitarily 
{\em in}equivalent quantum theories upon canonical quantization.
This arises from a remarkable ambiguity, not previously investigated,
in the construction of the classical (and hence the
quantized) Hamiltonian or Lagrangian.  This ambiguity is
illustrated for systems with one degree of freedom:
An arbitrary function of the
constants of motion can be introduced into this construction.  For
example, the nonrelativistic and relativistic free particles follow
identical classical trajectories, but the Hamiltonians or Lagrangians,
and the canonically quantized versions of these descriptions, are 
inequivalent.  Inequivalent descriptions of other systems, 
such as the harmonic oscillator, are also readily obtained.

\end{abstract}



\newpage

\section{Introduction}
\label{sect:intro}

Although there are quantum systems which cannot be obtained by
quantizing a classical system, the route through a classical
theory, via canonical quantization or
path integrals, is by far the most important way of
constructing a quantum theory. One of the most profound problems in
physics today is to quantize the classical theory of relativity, an
issue to which enormous and diverse efforts have been devoted
in the past 50 years.  It may therefore be useful to re-examine this
quantization procedure with care.  Here we discuss one interesting point
discovered in such a re-examination, although its implications (if any)
specifically for the quantization of gravity are as yet unclear. 
Rather, it appears to bear on the matter of quantizing classical systems
in general.

Given a classical system, we can directly observe
the particle trajectories.
Thus we can determine the
classical equations of motion, but not directly the Hamiltonian
or the Lagrangian.  From the equations of motion, we (i)  construct
the Hamiltonian, Lagrangian or action, which 
(ii) by canonical quantization
or a path integral, gives us the
quantum description.  The quantum mechanics of the system
is determined by the action, but the classical paths picked out by the
equations of motion concern only the extrema; one certainly
does not expect, in general, a functional to be uniquely
determined from its extrema.  From this point of view,
it is remarkable that the
quantum mechanics can be determined to a large extent
from the classical equations of motion.  The issue we address in this paper
is: {\em To what extent?}

There are well-known ambiguities in both steps (i)
and (ii) of this procedure.
In step (i), different Hamiltonians or Lagrangians
can be constructed from the
same equations of motion.  A famous example is the Aharonov-Bohm
effect \cite{ab}.  For an electron with charge $-e$ 
which is kept from entering a
long solenoid, the classical equation of motion is the same 
whether the magnetic field inside the solenoid is
turned on or not --- but the Hamiltonians (or Lagrangians)
are different.
In the absence of a magnetic field,
outside the solenoid the Hamiltonian is

\beq
H = \frac{{\bf p}^ 2}{2m} \quad ,
\eeql{eq:h1}

\nid
while with the magnetic field turned on, it is

\beq
H = \frac{ ( {\bf p} + e {\bf A} )^ 2} {2m} \quad .
\eeql{eq:h2}

\nid
Although classically this difference in the Hamiltonians is not
observable, it leads to observable effects upon quantization.

In step (ii) of the quantization procedure, we have the well-known
ordering ambiguity, which appears in various guises.  
In canonical quantization, it appears as different ways of
ordering products of the operators ${\hat x}$ and ${\hat p}$;
different orderings are possible as long as they lead 
to Hermitian operators with the same classical
limit \cite{fn2}.  In the configuration-space path-integral formulation,
it appears as the ambiguity in choosing the measure in the space
of paths, while in the phase-space path integral it appears as the
ambiguity in skeletonization \cite{hartle1}.  Corresponding to these
different manifestations are various proposals to fix this 
ambiguity \cite{hall}.

In this paper we address an ambiguity in step (i) of the
procedure, similar to but not the same as that of the
Aharonov-Bohm effect.  It is similar in that different Hamiltonians
or Lagrangians can be constructed 
for the same classical equations of motion.  However, in the case
of the Aharonov-Bohm effect the Hamiltonians (\ref{eq:h1}) and (\ref{eq:h2}) are
related by a canonical transformation

\beq
{\bf  x} \mapsto {\bf x}' = {\bf x} \quad,
\eeql{eq:canx}

\beq
{\bf p} \mapsto {\bf p}' = {\bf p} - e {\bf A} ( {\bf x} ) \quad ,
\eeql{eq:canp}

\nid
with a generating function of the second type \cite{goldstein},

\beq
F_2 = {\bf p}' \cdot {\bf  x} 
+ e \int {\bf  A} ( {\bf  x} ) \cdot d{\bf x} \quad .
\eeql{eq:genfn}

\nid
With (\ref{eq:canx}) -- (\ref{eq:genfn})
we can readily obtain \cite{goldstein} the
Hamiltonian (\ref{eq:h2}) from
form (\ref{eq:h1}).  Since the transformation is canonical, Poisson brackets are
preserved \cite{goldstein2}.  With the Poisson brackets turned into
commutators upon quantization \cite{fn2}, it is guaranteed that the two
Hamiltonians lead quantum-mechanically to the same local physics, although
globally they are different, as manifested in the Aharonov-Bohm effect.
In comparison, the ambiguity we examine here is that, in general, for
the same classical motions, there
exist different Hamiltonians which are {\em not} canonically
equivalent (even locally); these will
lead to different quantum behaviors.
It is amazing that such an ambiguity, which could have been
discovered 70 years ago, has not to our knowledge previously appeared in
the literature.

In Section \ref{sect:free} we describe the construction of different
Hamiltonians from a given equation of motion,
for systems with one degree of freedom.  It is shown
that the difference cannot be eliminated, even locally,
by a canonical transformation.  Section \ref{sect:ex} gives simple
examples, demonstrating that the different
Hamiltonians lead to different quantum descriptions.
Section \ref{sect:lag} gives the Lagrangian treatment,
in which the result can in many cases be relatively simple.
Section \ref{sect:dis} is a brief conclusion, including remarks
on coupled systems with more than one degree of freedom.  Where needed,
units with $\hbar = c = 1$ are used.

\section{Freedom in constructing the Hamiltonian}
\label{sect:free}

Suppose the equations of motion of a system
with one degree of freedom are found experimentally to be

\beq
\dot{p} + \frac{\partial H}{\partial x} = 0 \quad ,
\eeql{eq:eqp}

\beq
\dot{x} - \frac{\partial H}{\partial p} = 0 \quad ,
\eeql{eq:eqx}

\nid
for a certain function $H =H(p,x)$.  No doubt one can identify
this function $H$ as the Hamiltonian, and base canonical
quantization on it.  However, classically (2.1) and (2.2) are
indistinguishable from

\beq
g_1 (p,x) \left[ \dot{p}+ \frac{\partial H}{\partial x} \right] +
g_2 (p,x) \left[ \dot{x}- \frac{\partial H}{\partial p} \right]
= 0 \quad ,
\eeql{eq:neweoma}

\beq
g_3 (p,x) \left[ \dot{p}+ \frac{\partial H}{\partial x} \right] +
g_4 (p,x) \left[ \dot{x}- \frac{\partial H}{\partial p} \right]
= 0 \quad ,
\eeql{eq:neweomb}

\nid
as long as

\beq
\Delta \equiv g_1 g_4 - g_2 g_3 \ne 0 
\eeql{eq:det}

\nid
obtains.  In general, we can form not just linear combinations of 
(\ref{eq:eqp}) 
and (\ref{eq:eqx}), and the $g$'s can depend on $p$,
$x$ and the time $t$.  For simplicity, we shall restrict to
linear combinations in (\ref{eq:eqp}) 
and (\ref{eq:eqx}), and time-independent $g$'s.  This will suffice
to demonstrate the
ambiguity in constructing the Hamiltonian.

We examine the possibility that (\ref{eq:neweoma})
and (\ref{eq:neweomb}) are the Hamiltonian equations
of motion for another Hamiltonian $H'(p',x)$,
where the coordinate $x$ (being directly observable)
is unchanged, i.e., $x' = x$, but the momentum $p'$ could be different
from $p$.  This then requires

\beq
\dot{p}' + \left. \frac{\partial H'}{\partial x} \right|_{p'}
= g_1 \left[ \dot{p} + \frac{\partial H}{\partial x} \right]
+ g_2 \left[ \dot{x}-  \frac{\partial H}{\partial p} \right] \quad ,
\eeql{eq:neweleq1}

\beq
\dot{x} - \left. \frac{\partial H'}{\partial p'} \right|_{x}
= g_3 \left[ \dot{p} + \frac{\partial H}{\partial x} \right]
+ g_4 \left[ \dot{x}- \frac{\partial H}{\partial p} \right] \quad .
\eeql{eq:neweleq2}

\nid
The partial derivatives on the RHS of (\ref{eq:neweleq1}) and
(\ref{eq:neweleq2}) are understood to be 
at fixed $p$ or $x$.  By taking

\beq
p' = p' (p,x) \quad ,
\eeql{eq:newp}

\beq
H' = H' ( p'(p,x), x) \quad ,
\eeql{eq:newh}

\nid
the LHS of (\ref{eq:neweleq1}) and (\ref{eq:neweleq2}) can be rewritten
as functions of $x, \dot{x}, p$ and $\dot{p}$, as can the RHS.
Since $x, \dot{x}, p$ and $\dot{p}$ are independent, the
coefficients of $\dot{x}$ and $\dot{p}$ in (\ref{eq:neweleq1}) must separately
match.  By converting the partial derivatives on the LHS
from fixed $p'$ to fixed $p$ and $x$, we find

\beq
\frac{\partial p'}{\partial x} = g_2 (p,x) \quad ,
\eeql{eq:tr1}

\beq
\frac{\partial p'}{\partial p} = g_1 (p,x) \quad .
\eeql{eq:tr2}

\nid
The other terms in (\ref{eq:neweleq1}) give

\beq
\frac{\partial H'}{\partial x} 
- \frac{\partial p'  / \partial x} {\partial p' / \partial p} 
  \frac{\partial H'}{\partial p} 
=  g_1 \frac{\partial H}{\partial x} 
-g_2  \frac{\partial H}{\partial p} \quad .
\eeql{eq:tr3}

\nid
Likewise, (\ref{eq:neweleq2}) leads to

\beq
g_4 ( p,x) = 1 \quad ,
\eeql{eq:tr4}

\beq
g_3 ( p,x) = 0 \quad ,
\eeql{eq:tr5}

\beq
\frac{1}{\partial p' / \partial p} 
\frac{\partial H'} {\partial p} 
= \frac{\partial H}{\partial p} \quad .
\eeql{eq:tr6}

\nid
The conditions (\ref{eq:tr1}) -- (\ref{eq:tr6}) imply the pair of equations

\beq
\frac{\partial H'}{\partial x}
= g_1 (p,x) \frac{\partial H}{\partial x} \quad ,
\eeql{eq:newh1}

\beq
\frac{\partial H'}{\partial p}
= g_1 (p,x)  \frac{\partial H}{\partial p} \quad .
\eeql{eq:newh2}

\nid
Hence the integrability condition for $H'$ is

\beq
\frac{\partial H}{\partial x}  
\frac{\partial g_1}{\partial p}
 - \frac{\partial H}{\partial p}
\frac{\partial g_1}{\partial x}
\equiv \left[ H , g_1 \right]_{x,p} = 0 \quad ,
\eeql{eq:intcond}

\nid
with $[ \;\; ]_{x,p}$ being the Poisson bracket with respect to the
independent variables $x$ and $p$.  This means that $g_1(p,x)$ is
a constant of motion.

In one dimension, there is only one nontrivial constant
of motion, namely $H$.  Thus, we must have

\beq
g_1 = F(H) \equiv \frac{d {\tilde F}(H)}{dH} \quad .
\eeql{eq:defnF}

\nid
From (\ref{eq:newh1}) and (\ref{eq:newh2}),
$\partial H' / \partial x = 
\partial {\tilde F} / \partial x$, and
$\partial H' / \partial p = 
\partial {\tilde F} / \partial p$,
so that we get immediately

\beq
H' = {\tilde F}(H) \quad ,
\eeql{eq:newhint}

\nid
up to an irrelevant additive constant.

This expresses $H'$ in terms of $(p,x)$.  To express
$H'$ in terms of $(p', x)$, we note that (\ref{eq:tr2})
determines $p'$ as 

\beq
p' = \int_0^p dp \, g_1(p,x) \,+\, S(x) \quad ,
\eeql{eq:intnewp}

\nid
up to the arbitrary function $S(x)$.  In view of
(\ref{eq:tr1}), this is the only remaining freedom
in $g_2$.  In many examples, $H$ 
is even in $p$; if we want to maintain the same
condition for $H'$ in terms of $p'$, then we would
have to choose $S(x) =0$.

To be more explicit, the formal solution can be expressed
as follows.  We write

\beq
F(H) = \sum_n a_n(x) p^{2n} \quad .
\eeql{eq:powerF}

\nid
(For simplicity we take only even powers, though this
restriction is easily lifted.)  This yields 

\beq
p' = \sum_n \frac{1}{2n+1} a_n(x) p^{2n+1} \quad ,
\eeql{eq:newp2}

\nid
where, consistent with the assumption
that $H$ is even in $p$, we have set the
integration constant to $S(x) =0$.
It is easily checked that (\ref{eq:newp2}) solves (\ref{eq:tr1})
as well, because we have guaranteed the integrability
condition.  The inverse transformation is formally

\beq
p = \sum_n b_n(x) p'^{2n+1} \quad ,
\eeql{eq:newpinv}

\nid
with

\bea
b_0 &=& \frac{1}{a_0} \quad ,
\nonumber \\
b_1 &=& - \frac{1}{3} \frac{a_1}{a_0^4} \quad ,
\nonumber \\
b_2 &=& \frac{1}{3} \frac{a_1^2}{a_0^7}
- \frac{1}{5} \frac{a_2}{a_0^6} \quad ,
\eeal{eq:b}

\nid
etc.

Thus, given any $H$, we can obtain
a host of other Hamiltonians by choosing (i) 
an arbitrary fucntion $F(H)$ 
and (ii) an arbitrary
function $S(x)$.
The resulting
Hamiltonian equations are guaranteed to be equivalent to the
original ones, (\ref{eq:eqp}) and (\ref{eq:eqx}),
as long as $F$ is nonzero 
[cf. Eqs. (\ref{eq:det}), (\ref{eq:tr4}) and 
(\ref{eq:tr5})].

Finally consider
the Jacobian of the transformation
$(x, p) \mapsto (x' , p')$, where $x' = x$:

\beq
J = \det 
\pmatrix {
\partial x' / \partial x  &
\partial x' / \partial p \cr 
\partial p' / \partial x &
\partial p' \partial p  }
= [ x' , p' ]_{x,p}
= g_1(p,x) \quad .
\eeql{eq:jac}

\nid
A canonical transformation must have $J = 1$.  The so-called
extended canonical transformation \cite{goldstein} (with scale
transformation included) has $J = \mbox{constant}$, 
but not a function
of $p$ or $x$.  Provided $g_1 = F(H)$ is not chosen to
be a constant over all of phase space, the transformation
$(x,p) \mapsto (x', p')$ is not canonical.

\section{Examples}
\label{sect:ex}

\subsection{Free particle}
\label{subsect:fp}

A Newtonian free particle is described by the Hamiltonian

\beq
H = \frac{p^2}{2m} \quad .
\eeql{eq:hfree}

\nid
The choice

\beq
F(H) = \left( 1 - 2H/m \right)^{-3/2}
= \left( 1 - p^2/m^2 \right)^{-3/2} \quad ,
\eeql{eq:fex1}

\nid
leads to

\beq
H' = {\tilde F} = \left(m^2 - p^2 \right)^{-1/2} \quad ,
\eeql{eq:newhex1c}

\beq
p' = \frac{p}{ (1 - p^2 /m^2 )^{1/2} } \quad .
\eeql{eq:newpex1}

\nid
In terms of $p'$, we have

\beq
H' = \left( p'^2 + m^2 \right)^{1/2} \quad ,
\eeql{eq:newhex1b}

\nid
which is the Hamiltonian of a {\em relativistic}
free particle!  It is physically obvious
that by observing classical free particles [provided
the particles do not travel at speeds greater than unity, i.e.,
provided $p<m$, as per (\ref{eq:newhex1c})],
the two theories cannot be distinguished.
Although 
this relationship between the Newtonian and the relativistic 
free particle might not have a deep physical meaning,
this example clearly demonstrates that drastically different-looking theories
can correspond to the same classical motions.
It is well known that the quantum theory of a
relativistic point particle is very different from that of a
Newtonian free particle, with the former sharing some common
features with quantum gravity \cite{hartle1}.

Although the Hamiltonian (\ref{eq:newhex1b}) is nonpolynomial 
in $p$, and is therefore 
a non-local operator upon quantization, there is no reason why
canonical quantization cannot be based on it.  This has been
studied by Newton and Wigner \cite{newton} (see also Hartle and
Kuchar \cite{hartle1}).
The result can be stated simply: a Fourier
component with wave number~$k$ evolves in time~$t$ with a phase
$\sqrt{k^2 + m^2}\, t$.  This leads to the propagator
function $G'$ for $H'$, viz.,

\bea
G'(x,t;x_0 , t_0 ) 
&=& \int_{-\infty}^{\infty} \frac{dk}{2 \pi} \,
e^{ik(x-x_0)} 
e^{-i(k^2 + m^2 )^{1/2} (t-t_0 )} 
\nonumber \\
&=& \lim_{\epsilon \rightarrow 0^+ }
\left[ \frac{im(t-t_0 -i \epsilon) }{\pi \Delta \lambda} 
K_1 (m \Delta \lambda )
\right] \quad ,
\eeal{eq:green1}

\nid
with $\Delta \lambda = [(x-x_0)^2 - (t- t_0 -i\epsilon )^2 ]^{1/2}$,
and $K_1$ the usual modified Bessel function \cite{redmount}.  This
$G'$ is the probability amplitude for the wavefunction originally
localized at $x_0$ at time $t_0$ to be localized at $x$ at the later
time $t$, and has the same physical meaning as the familiar Newtonian
free-particle propagator $G$:

\beq
G (x,t; x_0 , t_0 ) 
= \sqrt { \frac{m}{2i \pi (t- t_0 )} }
\exp \left[ \frac{i m(x-x_0)^2}{2(t-t_0 )} \right] \quad ,
\eeql{eq:green2}

\nid
corresponding to the Hamiltonian (\ref{eq:hfree}).
So although
both the Hamiltonians (\ref{eq:hfree}) and (\ref{eq:newhex1b})
give straight lines as
classical trajectories, the quantum theories they yield via canonical
quantization are very different.

\subsection{A system with a non-trivial potential}
\label{subsect:pot}

Consider a one-dimensional system with the classical Hamiltonian

\beq
H = x^2 p^2 + x^3 p + \quarter x^4 \quad ,
\eeql{eq:hex2}

\nid
and take

\beq
F(H)= H^{-1/2} =
 \left(xp + \half x^2 \right)^{-1} \quad .
\eeql{eq:g1ex2}

\nid
The conditions (\ref{eq:tr1}) and (\ref{eq:tr2}) then give

\beq
p' = \frac{1}{x} \ln (xp + \half x^2 ) + S(x) \quad ,
\eeql{eq:newpex2}

\nid
for some function $S(x)$.  We choose, again for simplicity, 
$S(x)=0$.  The new Hamiltonian is

\bea
H' &=& 2H^{1/2} = 2 \left( xp + \half x^2 \right)
\nonumber \\
& =& 2 e^{x p'} \quad .
\eeal{eq:newhex2a}

The easiest way to analyse the properties of this Hamiltonian is
to perform an additional canonical transformation

\beq
{\tilde p}= x p' \quad , \quad {\tilde x} = \ln x \quad .
\eeql{eq:canex2}

\nid
The generating function is of the second type \cite{goldstein}:

\beq
F_2(x, {\tilde p} ) = {\tilde p} \ln x \quad .
\eeql{eq:genfnex2}

\nid
The Hamiltonian (\ref{eq:newhex2a}) becomes

\beq
{\tilde H} = 2 e^{{\tilde p}} \quad .
\eeql{eq:newhex2b}

\nid
The Hamiltonian equations of motion are then trivial, yielding the classical
paths

\beq
{\tilde x} (t) = 2 e^{{\tilde p}} t + \mbox{const} \quad ,
\eeql{eq:xcex2a}

\nid
which, via (\ref{eq:canex2}), gives

\beq
x(t) = e^{C_1 t + C_2} \quad ,
\eeql{eq:xcex2b}

\nid
where $C_1$ and $C_2$ are the two integration constants.
It can be checked that trajectory (\ref{eq:xcex2b})
indeed solves the Hamiltonian
equations of the original Hamiltonian (\ref{eq:hex2}). 

The canonical quantization of the Hamiltonian (\ref{eq:hex2})
yields
a different quantum theory than that of (\ref{eq:newhex2a})
[which is
the same as that of (\ref{eq:newhex2b})]; this
can be seen by comparing the Poisson
brackets in the two cases.  For any pair of functions

\bea
u &=& u( p', x) = u\Bigl( p' (p,x),x\Bigr) \quad ,
\label{eq:u}\\
v &=& v( p', x) = v\Bigl( p' (p,x),x\Bigr) \quad ,
\eeal{eq:v}

\nid
it is straightforward to verify

\bea
\left[u , v \right]_{x, p'}
&\equiv& \frac{\partial u}{\partial x}
\frac{\partial v}{\partial p'}
- \frac{\partial u}{\partial p'} 
\frac{\partial v}{\partial x}
\nonumber \\
&=& e^{x p'}
\left( \frac{\partial u}{\partial x} \frac{\partial v}{\partial p}
- \frac{\partial u}{\partial p} \frac{\partial v}{\partial x} 
\right) = e^{xp'} \left[ u , v \right]_{x,p} \quad ,
\eeal{eq:uv}

\nid
where the factor $e^{xp'}$ is just the inverse of
the Jacobian $g_1$ of the transformation.  As the Poisson
bracket gives the leading term in order of $\hbar$ of the quantum
commutator, the Hamiltonians (\ref{eq:hex2}) and (\ref{eq:newhex2a})
must correspond to
different quantum theories.  In particular, since the commutators
are different, the eigenvalue spectra for observables will be
different in the two theories.

\subsection{The harmonic oscillator}
\label{subsect:ho}

Even thoroughly understood systems can be subjected to this type of
transformation.  For the harmonic oscillator, experimental
evidence implies that the Hamiltonian must essentially be

\beq
H = \frac{p^2}{2} + \frac{x^2}{2} \quad .
\eeql{eq:hex3}

\nid
We consider three examples of transformations.

\nid
{\em Example 1}

\nid
We take

\beq
F = 1 + \frac{\epsilon}{2H}
= 1 + \epsilon \left( p^2 + x^2 \right)^{-1} \quad ,
\eeql{eq:defFex3}

\nid
with $\epsilon \ll 1$.  Integrating this in $H$, we find

\beq
{\tilde F} = H + \frac{\epsilon}{2} \ln H \quad .
\eeql{eq:newhex3c}

\nid
Using (\ref{eq:tr2}),
we find

\beq
p' = p + \frac{\epsilon}{x} \arctan \frac{p}{x} \quad .
\eeql{eq:newpex3b}

\nid
If $H'$ is now expressed in terms of $p'$ and $x$, we
get

\beq
H' = \frac{p'^2}{2} + \frac{x^2}{2}
+ \epsilon \left[ \frac{1}{2} \ln \frac{p'^2+x^2}{2} - 
\frac{p'}{x} \arctan \frac{p'}{x} \right] + O(\epsilon^2) \quad .
\eeql{eq:newhex3e}

\nid
{\em Example 2}

\nid
As another example, we choose

\beq
F = 1 + \epsilon (2H) = 1 + \epsilon(p^2 + x^2) \quad .
\eeql{eq:fex3b}

\nid
Follwing the same steps, we find

\beq
H' = \frac{p'^2}{2}
 + \frac{x^2}{2} 
 + \epsilon \left[ \quarter ( p'^2 + x^2 )^2 
 - \sfrac{1}{3}p'^4 
 - x^2 p'^2 \right] \quad ,
\eeql{eq:newhex3d}

\beq
p' = p + \epsilon 
\left( \sfrac{1}{3}p^3 + x^2 p \right) \quad .
\eeql{eq:newpex3d}

\nid
{\em Example 3}

\nid
In this case we take
$F = H$ 
and ${\tilde F} = \half H^2$.
The relation between $p'$ and $p$ is given by (\ref{eq:tr2}),
leading to

\beq
p' = \sfrac{1}{6}p^3 + \half x^2 p \quad ,
\eeql{eq:newpex3h}

\nid
where we have set the integration constant to zero
under the assumption that parity is maintained
(i.e., when $p \mapsto -p$, then $p' \mapsto -p'$).
The inverse relation cannot be expressed in closed
form, but is formally given by the power series

\beq
p = \frac{2}{x^2} p' - \frac{8}{9x^8} p'^3 + \cdots \quad .
\eeql{eq:pex3inv}

When (\ref{eq:pex3inv}) is substituted into 
$H' = {\tilde F} = (p^2 + x^2 )^2/8$, we obtain $H' = H'(p',x)$, which
is guaranteed to give the same classical motion.
However, this is a complicated expression.
We shall see later that the complication arises
entirely from $p'$, and the same situation has a simple
description in the Lagrangian formulation because
$p'$ will not appear.

\section{Lagrangian formulation}
\label{sect:lag}

The preceding examples show that at least part of the
complication arises from the transformation from $p$ to
$p'$, and the inverse transformation, e.g.,
(\ref{eq:pex3inv}).  This would suggest that the formalism
may be simpler in a Lagrangian description, in so far
as neither $p$ nor $p'$ need appear.

\subsection{Formalism}
\label{subsect:lagform}

The Lagranian equation of motion is

\beq
\frac{d}{dt} \left( \frac{\partial L}{\partial \dot{x}} \right)
- \frac{\partial L}{\partial x} = 0 \quad .
\eeql{eq:lag1}

\nid
with $L = L(x, \dot{x})$.  In greater detail, and adopting the
notation $( x, \dot{x} , \ddot{x} ) \mapsto (x, y, z)$,
and $L_x = \partial L / \partial x$ etc., we have

\beq
z L_{yy} + y L_{xy} - L_x = 0 \quad .
\eeql{eq:lag2}

\nid
For this to be a non-degenerate second-order
differential equation, we shall assume $L_{yy} \ne 0$.
(In the Newtonian case, $L_{yy} = m$.)

This could just as well be written with an extra multiplcative
factor of $F = F(x,y)$.  If this modified equaton is to
be the Euler-Lagrange equation derived from another Lagrangian
$L'$, then we must have

\beq
z L'_{yy} + y L'_{xy} - L'_x 
= F \left( z L_{yy} + y L_{xy} - L_x \right) \quad .
\eeql{eq:lag3}

\nid
We have used the notation $F$ in anticipation 
that this function will turn out to be the same
as $g_1$ in the Hamiltonian formulation.

In the above equation, $z$ appears only where shown explicitly,
and since (\ref{eq:lag3}) must hold as an identity, the terms
with and without $z$ have to be separately equal, leading
to two conditions

\beq
L'_{yy} = F L_{yy} \quad ,
\eeql{eq:lag4}

\beq
yL'_{xy} - L'_x = F \left( yL_{xy} - L_x \right) \quad .
\eeql{eq:lag5}

In these expressions, we write
the quantities on the LHS as

\bea
L'_{yy} &=& \partial_y L'_y = \partial_y p' \quad , 
\nonumber \\
L'_{xy} &=& \partial_x L'_y = \partial_x p' \quad .
\eeal{eq:lag8}

\nid
Their compatibility requires the integrability
condition that $\partial_x \partial_y p'
= \partial_y \partial_x p'$.  Some arithmetic then
leads to

\beq
y F_x L_{yy} - y F_y L_{xy} + F_y L_x = 0 \quad .
\eeql{eq:lag9}

On the other hand, we can consider the quantity

\bea
&&L_{yy} \dot{F}
\nonumber \\
&=&L_{yy} \left( F_x \dot{x} + F_y \dot{y} \right)
\nonumber \\
&=&L_{yy} \left( F_x y + F_y z \right) \quad ,
\eeal{eq:lag10}

\nid
and use the equation of motion (\ref{eq:lag2})
to eliminate $z L_{yy} $.  This shows that
$L_{yy} \dot{F}$ is exactly the LHS of (\ref{eq:lag9}),
implying $\dot{F}=0$, i.e., $F$ is
a constant of motion.

In fact, it is easy to identify
what $F$ is.  Note that in (\ref{eq:lag5}), we can write
the bracket as

\beq
yL_{xy} - L_x 
= \partial_x \left( yL_y - L \right)
= \partial_x H \quad ,
\eeql{eq:lag6}

\nid
given $L_y = p$.  The same holds for the analogous
expression involving $L'$.  Thus (\ref{eq:lag5})
is equivalent to
$\partial_x H' = F \partial_x H$,
confirming that $F$ is indeed the same as $g_1$ introduced
previously [cf. (\ref{eq:newh1})].  However, in the rest of this Section, we
shall not rely on any results from the Hamiltonian treatment.

Thus, once we choose any constant of motion $F$, we can
integrate to find $L'$, up to a function $S(x)$.

\subsection{Free particle}
\label{subsect:lagfree}

We return to the free particle,
with
$L = \half my^2$,
$L_y = my$,
$L_{yy} = m$.
We choose $F = ( 1- y^2)^{-3/2}$, which
is a constant of motion, and solve for

\beq
L'_{yy} = F L_{yy} = m ( 1- y^2)^{-3/2} \quad ,
\eeql{eq:lag12}

\nid
giving, up to integration constants which we set to zero,

\beq
L' = - m ( 1- y^2 )^{1/2} \quad ,
\eeql{eq:lag13}

\nid
which is the Lagrangian for the relativistic
free particle.

\subsection{Harmonic oscillator}
\label{subsect:lagsho}

For the harmonic oscillator, we have 
$L = \half y^2 - \half x^2$,
$L_y = y$, 
and $L_{yy} = 1$.
We choose 
$F = \half y^2 + \half x^2$,
which is a constant.  Thus, $L'$ is to be found
by integrating

\beq
L'_{yy} = F L_{yy} = \half y^2 + \half x^2 \quad ,
\eeql{eq:lag16}

\nid
leading to

\beq
L' = \sfrac{1}{24} y^4 + \sfrac{1}{4} y^2 x^2 + f(x) \quad .
\eeql{eq:lag17}

\nid
There should be two integration constants, but we have
set the term linear in $y$ to zero using the assumption
of parity.  The function $f(x)$ is easily determined
by considering the terms without any $y$ in (\ref{eq:lag5}):

\beq
-f'(x) = \half x^2 \left( +x \right) \quad .
\eeql{eq:lag18}

\nid
Thus we have

\beq
L' = \sfrac{1}{24} \dot{x}^4 + 
\sfrac{1}{4} \dot{x}^2 x^2 - \sfrac{1}{8} x^4 \quad .
\eeql{eq:lag19}

\nid
It can be verified directly, without any of the
general formalism given above, that this Lagrangian
gives the same classical motion, with the Euler-Lagrange
equation being the usual one multiplied
by $F$.

In fact, this is the same as the last example presented
under the Hamiltonian treatment of the simple harmonic
oscillator, but now expressed in closed form.  
Comparison shows that the Lagrangian
formulation is much simpler is this case, since
all the complications lie in $p' = L'_y$.

\subsection{Path integral}
\label{subsect:path}

In all these cases the quantum
mechanics derived from $L$ and $L'$ are different.
This is most easily seen by verifying that,
in the path-integral formulation,
the phases associated with paths are different
in nontrivial ways.
Consider two paths $x_a(t)$
and $x_b(t)$ with the same end-points at $t = t_1$
and $t=t_2$ [i.e., 
$x_a(t_1) = x_b(t_1) =x_1$, 
$x_a(t_2) = x_b(t_2) =x_2$], 
and the action integrals
$S_a = \int_{t_1}^{t_2} dt \, L[ x_a ]$ etc.
The differences
$\Delta S = S_a - S_b$ and
$\Delta S' = S'_a - S'_b$
are clearly different in general. 
Therefore the quantum
mechanics as defined by the path integral will
be different in the two cases.

\section{Discussion}
\label{sect:dis}

We have shown that just as in the case of Aharonov-Bohm effect,
different Hamiltonians can be constructed for the same classical
equation of motion.  However, unlike the case of the
Aharonov-Bohm effect, these Hamiltonians are {\em not} related by
canonical transformations.  Although they must be regarded as the same
classical theory --- since it is the trajectories or 
equations of motion, not the Hamiltonian, which are observed classically
--- these variant
Hamiltonians give different quantum theories upon canonical quantization.

We may place the ambiguity discussed here into context by
classifying the known quantization ambiguities into three
levels.  First, there are {\em small} differences in the
sense that $H$ and $H'$ agree classically but differ
by $O(\hbar)$ quantum-mechanically; the operator-ordering
ambiguity is of this type.  Second, there are cases where
$H - H'$ is not small (in the above sense), but can be made small
by performing a canonical transformation on one of them.
The canonical transformation may be global (in which case the
situation is the same as the first type), or only local,
as in the Aharonov-Bohm effect.  The third level
are differences between $H$ and $H'$
that cannot be made small by a canonical transformation.
The ambiguities discussed in this paper are of this type,
and in a sense are the most ``serious."

This possibility of {\em in}equivalent 
Hamiltonians raises several
interesting questions, to which we do not at present
have complete answers.
One might be tempted to impose the requirement that the Hamiltonian must be
polynomial in the momentum.  However, this seems to be rather 
{\em ad hoc} 
as a fundamental principle. Moreover, in the case of a free particle, 
nature favors the more complicated relativistic Hamiltonian 
over the simpler
form (\ref{eq:hfree}).  In the case of the harmonic oscillator, 
Hamiltonian\ (3.24),
which is quadratic in $p$, seems to be the preferred one. It is not
clear whether either of the alternatives
(\ref{eq:hex2}) or (\ref{eq:newhex2a}) is more
appropriate than the other.  
Is there any guiding principle in choosing
which one to use, in general?  
Are all the other possibilities truly forbidden?

The discussion in this paper, and all the examples,
refer to systems with only one degree of freedom. 
This may already be of some interest, since 
some models of mini-superspace contain only
one degree of freedom \cite{super}.

However, the interesting
question is, of course, whether these ambiguities extend
to systems with more degrees of freedom.  
The most general argument is that, for any number of degrees
of freedom, one has no right to expect a functional (the action)
to be determined by its extrema (the classically allowed
trajectories), and therefore the existence of alternate
Hamiltonians and Lagrangians should be the rule rather than
the exception.  The challenge will be to discover interesting
examples, and if possible to characterize and classify
all the possible ambiguities.

\acknowledgments
This problem may be regarded as one of {\em inversion}:
how a system can be determined from the observables.
The work of W.~M.~S. and K.~Y. on inversion is supported in part
by a grant (CUHK 4006/98P) from the Hong Kong Research Grants Council.




\end{document}